\documentstyle[aps,multicol,subeqn,epsf,rotate,citesort]{revtex}
\newcommand{\gl}[1]{Eq. (\ref{#1})} 
\newcommand{\gls}[2]{Eqs. (\ref{#1},\ref{#2})}

\def\gtrless{\raise2.5pt\hbox{$>$}\llap{\lower2.5pt\hbox{$<$}}}
\def\gtrapprox{\raise2.5pt\hbox{$>$}\llap{\lower2.5pt\hbox{$\approx$}}}
\def\lssapprox{\raise2.5pt\hbox{$<$}\llap{\lower2.5pt\hbox{$\approx$}}}
\newcommand{\bsq}[1]{\begin{subequations}\label{#1}}
\newcommand{\esq}{\end{subequations}}
\newcommand{\beq}[1]{\begin{equation}\label{#1}}
\newcommand{\eeq}{\end{equation}}
\newcommand{\beqa}[1]{\begin{eqnarray}\label{#1}}
\newcommand{\eeqa}{\end{eqnarray}}
\newcommand{\fur}{\qquad\mbox{for }\, }
\newcommand{\wer}{\qquad\mbox{where }\quad}
\begin{document}

\title{Aspects of the dynamics of colloidal suspensions:
Further results of the mode--coupling theory of structural relaxation}
\author{M.~Fuchs and M. R.~Mayr}
\address{Physik-Department, Technische Universit{\"a}t M{\"u}nchen, 85747
Garching, Germany}
\date{July 20, 1999}
\maketitle

\begin{abstract}
Results of the idealized mode--coupling theory for the structural relaxation in
suspensions of hard--sphere  colloidal particles are presented and
discussed with regard to recent light scattering experiments. 
The structural relaxation becomes non--diffusive for 
long times, contrary to the
expectation based on the de Gennes narrowing concept.
A  semi--quantitative connection of the wave vector dependences of the
relaxation times and amplitudes of the final
$\alpha$--relaxation  explains the approximate
scaling observed by Segr{\`e} and Pusey [Phys. Rev. Lett. {\bf 77}, 
771 (1996)]. 
Asymptotic expansions lead to a qualitative understanding of
density dependences in generalized Stokes--Einstein relations. This relation is
also generalized to non--zero frequencies thereby yielding support for a
reasoning  by Mason and Weitz [Phys. Rev. Lett
{\bf 74}, 1250 (1995)]. 
The dynamics transient to the structural relaxation  is discussed with
models incorporating short--time diffusion and  hydrodynamic
interactions for short times.
\bigskip

\medskip

\end{abstract}
\begin{multicols}{2}
 
\section{Introduction \label{sekt1}}

The dynamics of suspensions of colloidal particles has been the topic
of active research for many years \cite{Pusey91,russel}. Whereas the motion of
isolated Brownian particles has been well understood for long, less is 
known about the dynamics of concentrated suspensions. Direct particle 
interactions and solvent mediated hydrodynamic interactions (HI) are 
important if the 
colloidal volume packing fraction increases above a few percent \cite{Pusey91}.
Experimental studies mainly employing dynamic light scattering (DLS)
have provided a wealth of information on dense systems
and are also the stimulus for the theoretical work presented in this 
contribution. 

The pioneering
study by van Megen and coworkers of the liquid to glass
transition in hard--sphere like colloidal dispersions has provided
detailed data on the density fluctuations at this dynamic, 
ergodic--to--nonergodic transition
\cite{Megen91,Megen91b,Megen93,Megen93b,Megen94,Megen94b,Megen95,Megen98}.
Besides their intrinsic 
interest, these experiments also made possible  quantitative
tests
\cite{Megen91,Megen91b,Megen93,Megen93b,Megen94,Megen94b,Megen95,Megen98,Goetze91}
of predictions from the idealized mode--coupling theory (MCT)
\cite{Goetze91b,gs}. 
Agreement of experiment and theory within errorbars of ca.  15 \% has been
reported.  This comparison, which up to now has tested
leading order asymptotic  predictions and has thus restricted
the validity of the theoretical results to 
small separations from the critical density, provides support for the glass 
transition scenario as described by MCT, which has also been studied
for colloidal micronetwork spheres \cite{Bartsch93b,Bartsch95c}, charged
colloidal systems \cite{haertl95}, and colloidal emulsions \cite{gang}.
Recent DLS experiments by Segr{\`e}, 
Pusey and coworkers study hard sphere like systems at
lower colloidal densities and report unexpected and seemingly unrelated
scaling properties of the dynamic scattering functions
\cite{segre96,pusey97,segre97}.  
Thus the question arises for which density range below the glass transition
the MCT describes the dominant physical mechanism observed in the dynamics
of concentrated colloidal fluids and whether the reported scalings can be 
explained by MCT. Studies of a generalized Stokes--Einstein relation
\cite{Segre95,pusey97} and optical measurements by Mason and Weitz
\cite{mason95b} further  
raise the question about the connection of the viscoelastic moduli
\cite{Mason95} to the 
collective and self particle motion at rather high densities which can also 
be considered using MCT \cite{johanprl}. 

The MCT was developed starting from theories of the dynamics of 
simple liquids upon the realization that in that subsystem of the equations 
of motion which aimed at describing the structural relaxation  
there exists a bifurcation separating ergodic from
non--ergodic motion \cite{Leutheusser84,Bengtzelius84}.
 The physical mechanisms held responsible 
have been called ``cage effect'' and  ``back flow'' phenomenon
\cite{gs,Goetze95}. 
This transition was suggested as origin of
the slowing down and of the anomalies of the dynamics at the glass 
transition. The idealized MCT studies the structural relaxation neglecting
all other, possibly present, long--time dynamical effects
\cite{Goetze84,Goetze85,Goetze91b}.  The extended MCT
discusses long--time ergodicity restoring corrections
\cite{Das86,Goetze87,Fuchs92b,Goetze95}. 
 The bifurcation at critical values of the thermodynamic parameters 
like the colloidal packing fraction, $\phi$,  
introduces a small (separation) parameter, $\varepsilon=(\phi-\phi_c)/\phi_c$,
and the possibility of asymptotic expansions in $\varepsilon$; see  
Refs. \cite{Goetze91b,Franosch97,Fuchs98} for references and
detailed results.  

Two asymptotic scaling law regions can be shown. In the first,
for intermediate times, the feedback mechanism of caging of particles, 
causes an  ergodic--to--nonergodic bifurcation, which   is characterized by
universal power law decays. During the second, for longer
times, the collective rearrangements of the cages requires cooperative
dynamics, like the build up of back flow patterns first discussed for liquid
helium \cite{Feynmann56,Feynmann57,Goetze76}.  The strongly correlated dynamics
manifests itself in a coupling of the time scales for this final process of the
structural relaxation. As it describes, in the liquid, the decay of the
incipient frozen glassy structure, it is not surprising, that its MCT
description requires detailed information about the equilibrium structure.

As the leading--in--$\varepsilon$ asymptotic results exhibit numerous
nontrivial universal features, experimental
tests of MCT mainly address these and thus a number of corrections need to 
be considered:  $(i)$ The structural relaxation itself leads to corrections
of higher order in $\varepsilon$ which limit the range of validity of the
leading asymptotics. $(ii)$ The short--time or microscopic dynamics affects
the transient to the structural dynamics and needs to be considered if no 
clear separation of time scales is possible. $(iii)$ Long--time relaxational
mechanisms may be present, which bypass the structural relaxation, and lead
to faster decay. The third correction appears to be absent
in colloidal suspensions at the densities of interest for the present study
and thus shall be neglected in the following \cite{Megen95,Megen98}.
It is interesting to mention, though, that in colloidal emulsions 
 droplet shape fluctuations  cause long--time relaxation and
can be explained within the extended MCT \cite{gang}.  
Theoretical understanding of the first correction effect in lowest relevant 
order in $\varepsilon$ has been achieved recently \cite{Franosch97,Fuchs98} and
is the basis 
of the present considerations. Lacking a deeper understanding of the
microscopic transport effects of colloidal suspensions (point $(ii)$ above) 
a qualitative study shall be undertaken incorporating short--time effects
with the most simple approximations compatible with the MCT description of 
the structural relaxation. Thus the limit of the predominance of the 
structural relaxation is estimated from numerical solutions of the MCT 
equations using simple models for the microscopic transient dynamics of 
colloidal suspensions. Brownian short--time diffusion with
and without hydrodynamic  interactions is considered \cite{Pusey91}. 

The paper is arranged as follows. Section II summarizes the equations of
motion of the idealized MCT. Section III presents and discusses our results,
focusing first on the aspects purely structural--relaxational and then on
the microscopic transient effects. The mentioned experimental findings are 
addressed in section III.C. Short conclusions end the paper.

\section{Equations of motion \label{sekt2}}

The idealized MCT of the liquid to glass transition leads to a closed set of 
non--linear equations of motion for the density fluctuations
\cite{Leutheusser84,Bengtzelius84,Goetze91b}.  Other
dynamical 
variables are connected to them via the Zwanzig--Mori formalism and 
mode--coupling approximations. The theory aims at a description of the 
structural 
relaxation as it emerges from the (microscopic) short--time dynamics and 
slows 
down owing to the increasing density and thus increasing importance
of particle interactions. The 
central 
quantity, the (normalized) intermediate scattering function,
$\Phi_q(t)=\frac1N \langle \varrho_q^*(t) \varrho_q(0) \rangle/S_q$, 
measures the time and wave vector dependence of 
the collective density fluctuations, $\varrho_q(t)$,
around the average homogeneous density
which for hard--sphere particles of diameter $d$ is converted to the volume
packing fraction. The normalization is provided
by the static 
structure factor, $S_q=\langle|\varrho_q|^2\rangle/N$ \cite{hansen}.

The equations of motion of the idealized MCT 
 from which asymptotic analysis extracts the  physically 
relevant long--time dynamics can be summarized as \cite{gs}:
\beqa{e1}
& \Phi_q(t) -
 m_q(t) + \frac{d}{dt} \int_0^t dt'\; m_q(t-t')\, \Phi_q(t') = 0 \; , &\\
\label{e2}
& m_q(t) =  \sum_{{\bf k} + {\bf p} = {\bf q}} \; V(q;k,p)
\, \Phi_k(t)\, \Phi_p(t)\; , & 
\eeqa
where the coupling constants or vertices, $V$, are uniquely specified by 
the static  structure factor, $S_q$; see Refs. \cite{Sjoegren80,Bengtzelius84}
for  explicit  formulae.
Equation (\ref{e2}) approximates the autocorrelation function of the
fluctuating forces by considering a force to arise between two density
fluctuations interacting via an effective potential. Then the correlation
function of the four density fluctuations is approximated by the squared
density correlators and the effective potential enters the vertices.

In \gls{e1}{e2} the structural dynamics results from the equilibrium 
state of 
the fluid 
as captured in $S_q$ which is the only input.  Thus the long--time
structural  relaxation 
of a dense suspension of interacting Brownian particles is predicted to be 
identical to the one of an atomic system if the interaction potentials 
of both systems coincide. Here we will consider hard sphere potentials only.
By ansatz, MCT neglects the possibility of an
ordered, crystalline  state and thus in experimental comparisons 
crystallization
has to be prevented. Then the structure factor $S_q$ of a liquid of hard 
spheres
is known to be a regular function smoothly varying with
 packing fraction \cite{hansen} and consequently
the vertices in \gl{e2} are regular functions of the single (for a liquid of
hard spheres)  thermodynamic state  parameter $\phi$.  

Brownian particles diffuse on (appropriately chosen) 
short distances and thus for short times \cite{Pusey91}. 
Incorporating this into MCT 
leads to the following simple model of colloidal 
suspensions close to the glass transition \cite{Fuchs95,Franosch97}: 
\beq{e3}
 \Phi_q(t) -
 m_q(t) + \frac{d}{dt} \int_0^t dt'\; m_q(t-t')\, \Phi_q(t')  =
\frac{-1}{q^2 D_q^{\rm s.} }\dot{\Phi}_q(t) \; .
\eeq
This equation replaces \gl{e1} and extends it to short times where
the initial condition runs $\Phi_q(t)\dot{=}1-q^2 D_q^{\rm s.} t$. 

Two approximations for the
short--time diffusion coefficient, $D_q^{\rm s.}$, are widely used in 
theoretical work on colloidal dispersions
and differ in the treatment of the solvent effects. In the most
 simple
model of Brownian diffusion the interaction of the solvent with the 
colloidal particles is modeled with a single friction coefficient, 
$\zeta_0$ \cite{Pusey91}. 
This leads to $D_q^{\rm s. (B)}=D_0/S_q$, where $D_0$ is 
given by 
Einstein's law, $D_0=k_BT/\zeta_0$, and the denominator
 arises from particle
interactions as was first argued by de Gennes in a related context
\cite{degennes59}. 
This approximation is not satisfactory except for very low packing 
fractions, because the solvent also leads to 
long--ranged and quasi--instantaneous interactions of the colloidal 
particles,
called hydrodynamic interactions (HI). Whereas the HI do not affect
 the equilibrium
statistics and thus $S_q$ of the colloidal system, their effects on 
short time
scales cannot be neglected in general
and are captured in a wave vector dependent amplitude, $H_q$ \cite{Pusey91}:
$D_q^{\rm s. (HI)}=D_0 H_q/S_q$. Progress on a detailed theory for $H_q$
has proven very difficult but has culminated in accurate 
results for it up to intermediate packing fractions
\cite{beenakker83,beenakker84}. At the considered high packing
fractions however,  $D_q^{\rm s. (HI)}$ can only be estimated from experiments
or simulations at present  \cite{segre95b,segre96}.
 Note, that our approach to incorporate
HI into the MCT equations of motion only via $D^{\rm s.}_q$ differs
from the one developed by N{\"a}gele and others \cite{naegele97}, which aims at
describing the dynamics at lower packing  fractions,  and which would affect
the structural long--time dynamics. Our approach also differs
from the work of Cohen et al. who incorporate aspects of the cage effect into
an effective  short--time $D^{\rm s.}_q$ \cite{Cohen91,Cohen98}.
The role of the HI here also differs from a recent theory
of Tokuyama et al. \cite{toku1,toku2,toku3} who consider the HI in
nonequilibrated colloidal suspensions.

Equations (\ref{e2}) and (\ref{e3}) have been solved repeatedly without 
HI and
with different approximations for the structure factor of the 
hard--sphere fluid \cite{Fuchs92b,Fuchs95,Franosch97,Fuchs98};
for the details of the numerical calculations see the quoted references.
Various aspects of the known solutions will be connected to recent 
experimental
observations in this contribution and new solutions taking HI into 
account via 
$D^{\rm s. (HI)}_q$ will be presented, which are an extension of the
 calculations
in  Refs. \cite{Franosch97,Fuchs98}. 

Figure \ref{fig1} shows the short--time diffusion coefficients entering the 
numerical calculations discussed in the following. The short time 
diffusion coefficient without HI, $D^{\rm s. (B)}_q$, follows immediately
from the hard sphere structure factor, where the Percus--Yevick approximation
is used \cite{hansen}. The $S_q$ shown  also enter the vertices in \gl{e2}.
 The short--time diffusivity with HI, $D^{\rm s. (HI)}_q$, is chosen 
as shown in Fig. \ref{fig1}.  It is aimed at a discussion of the dynamics with
HI transient to the structural relaxation, and thus, for the
high densities considered, a rough approximation modeled from the experiments
in \cite{segre95b,segre96} is used. 
Outside the window $0.77\le q/q_p\le2.4$, $D^{\rm s. (HI)}_q$ is assumed
constant for simplicity. The values of $D^{\rm s. (HI)}_q$ for $q\to0$,
$q=q_p$ and $q\to\infty$ are adjusted to 0.2:1.74:1.0 mimicking the measured
ratios \cite{segre96,segre95b}. Within the mentioned wave vector window, 
the experimental data are modeled  by:
$D^0/D^{\rm s. (HI)}_q = -x(q/q_p)^2/\ln f^c_{q/q_p}$,
where $f^c_q$ is the MCT critical non--ergodicity parameter,
and  $x=0.29$ leads to a continous matching.
``De Gennes narrowing''
 is present in $D^{\rm s.}_q$ in both approximations, as 
its inverse varies in phase with the structure factor $S_q$ and an 
appreciable slowing down of the short--time dynamics for wave vectors $q$ 
around the principal peak at $q=q_p$ results. 

Some representative numerical results for the collective density correlators,
$\Phi_q(t)$, 
obtained as specified in \cite{Franosch97,Fuchs98}, 
are exhibited in Fig. \ref{fig2},  where also the 
shown wave vectors are indicated. The correlators of Fig. \ref{fig2}(a) 
correspond
to a density rather close to the critical liquid--to--glass bifurcation point,
$\phi_c=0.516$, of this model. The reduced distance equals 
$\varepsilon=(\phi-\phi_c)/\phi_c=- 10^{-n/3}$, with $n=9$, where, as in the 
following, in order to simplify comparison with
Refs. \cite{Franosch97,Fuchs98}, the packing fractions 
will be reported by stating the number $n$. Decreasing the  packing fraction
to $\varepsilon=-0.1$ corresponding to $n=3$, results in the  intermediate
scattering functions of Fig. \ref{fig2}(b). Results with and without HI as
shown in Fig. \ref{fig2}  for various packing fractions will be discussed in
the  following. 
Only the dynamics in the colloidal liquid phase is shown, where the
correlators decay to zero during the final relaxation process, because the
mentioned experimental studies focus on this final decay; for MCT results on
hard sphere glasses at higher packing fractions see Refs.
\cite{Fuchs95,Franosch97,Fuchs98}. It is also interesting to note that short
range attractions can increase the colloidal glass packing fraction appreciably
\cite{Fabbian99,Bergenholtz99}. 

\section{Results and discussion \label{sekt3}}

\subsection{Leading asymptotic scaling laws}\label{sekt3a}

In lowest order in the separation parameter $\varepsilon$, the MCT predicts the
existence of two divergent time scales with two different scaling laws
describing the dynamics in expanding windows in time or frequency; see the
Refs. \cite{Goetze84,Goetze85,Goetze87b,Goetze91b,Franosch97}
  for detailed derivations and reviews of these results.

In the first or $\beta$--scaling law window, a factorization property allows to
separate the sensitive and rather universal dependences of the dynamics on the
separation parameter and on time from the system specific dependences like 
spatial variation.
\beq{e4}
\Phi_q(t) = f^c_q + h_q\;  G(t,\varepsilon) \fur |\Phi_q(t)-f^c_q| \ll 1\; .
\eeq
The $\beta$--correlator is given by a homogeneous function, $G\propto
\sqrt{|\varepsilon|}\; g^\lambda_\pm(t/t_\varepsilon)$,  specified by one
system specific  parameter $\lambda$, which can be calculated for simple
liquids from $S_q$ and determines all exponents of MCT \cite{Goetze85}. 
The first divergent scaling time
$t_\varepsilon =t_0\; |\varepsilon|^{-(1/2a)}$, lies in the center of the
window of validity 
of \gl{e4} and, below the critical density, can be taken
from the root of $G$: $\Phi_q(t= t_\varepsilon)=f^c_q$.   The one parameter
$t_0$, the crossover or matching time, remains as only remnant of the
short--time or transient motion and can only be obtained from matching the
asymptotic results to the full dynamics including some short--time model.
In Fig. \ref{fig2}(a) one notices that $t_0$ differs by a factor $1.2$
 for the two
models of the transient. A shift of the curves with HI relative to the ones
without HI collapses both sets of curves for times $t\, \gtrapprox\,
 0.1 d^2/D_0$.  The great simplification of the
dynamics provided by the factorization in \gl{e4} may be interpreted as 
resulting from a localization transition close to which density fluctuations
relax via local rearrangements and not via  mass--transport over larger
distances. 
If the spatial variation of $f_q^c$ and $h_q$ is studied in detail
\cite{Barrat89},  the
localization may be traced back to the  ``cage--effect'', that particles are 
surrounded by next--neighbor shells whose ability to cage the central
particles depends on the fluctuations of the 
local structure and thus, in a cooperative manner, on the dynamics of the caged
particles themselves. 

In the second scaling law region,  another set of divergent time scales, called
$\alpha$--relaxation times $\tau_q$, appears and $\alpha$--master curves
describe the final relaxation of the density correlators from $f^c_q$ to zero
during that time window \cite{Goetze87b}.
\beq{e5} 
\Phi_q(t) \to \tilde{\Phi}_q(t/t'_\varepsilon) \fur \varepsilon\to0-
\;,\;\;  t/t'_\varepsilon=\mbox{fixed}\; .
\eeq
This superposition principle states that the final relaxation processes
(asymptotically)  depend on the distance to the critical point only via the
relaxation  times $\tau_q$, which moreover are coupled,
$\tau_q = \tilde{\tau}_q t'_\varepsilon$, and diverge upon approaching
$\phi_c$: $t'_\varepsilon=t_0 |\varepsilon|^{-\gamma}$ with $\gamma>(1/2a)$.
The equations, which the $\tilde{\Phi}_q(\tilde{t})$ obey, are obtained in a
special limit from \gls{e1}{e2} and are consequently independent of the
microscopic short--time dynamics \cite{Goetze91b,Franosch98}.
The resulting two--step relaxation scenario of the idealized MCT has been fully
worked out for some simple liquids;
see Refs. \cite{Bengtzelius84,Barrat89,Fuchs92b} 
for calculations of the exponents and master functions of
the two scaling regions for a hard sphere liquid.

The quantities of most immediate interest to experimental observations of the
final or $\alpha$--relaxation process are the relaxation times \cite{taudef1}.
Figure \ref{fig4} presents results for the asymptotic $\tau_q$ from the model
specified above  and compares them to previous
calculations using a different approximation (Verlet--Weis form)
for the static structure factors
of a hard sphere liquid \cite{Fuchs92b}. 
Very small differences in the $\tau_q$ result from the two 
approximations to $S_q$.
A discussion of short--time sum rules for colloidal suspensions as done in the
spirit of de Gennes \cite{degennes59} leads to the prediction of (short--time)
relaxation times obeying $\tau_q^{\rm s.} = (1/q^2 D^{\rm s.}_q)$. 
Such a behavior for the Brownian model, scaled to match $\tau_q$ for $q=q_p$,
is also indicated  in Fig. \ref{fig4}. The MCT $\alpha$--relaxation times 
obtained from \gls{e1}{e2}, where the transient does not enter,
 and  the results from the  short--time sum rules
qualitatively are similar for not--too--small wave vectors because both 
vary in phase with the structure factor. 
Their different physical origins, however, clearly show up for small wave
vectors where the short--time relaxation times,  $\tau_q^{\rm s.}$,
become diffusive, whereas the MCT $\alpha$--relaxation times $\tau_q$  become
wave vector independent as first anticipated in 
Mountain's description of Brillouin scattering in supercooled atomic liquids
\cite{Mountain66}. 
Although the collective density fluctuations of the colloidal
Brownian particles are diffusive on short time scales due to random collisions
with solvent molecules, during the structural relaxation
only stress fields arising from colloid--colloid particle
interactions survive out to long times. Thus large distance density
fluctuations decay by local particle rearrangements. 
The strong slowing down of
$\tilde{\Phi}_q(\tilde{t})$ on length scales of the order of the average
next--neighbor distance indicates local and cooperative particle
rearrangements  and is reminiscent of the back flow phenomenon familiar
from simple liquids \cite{Feynmann56,Feynmann57,Goetze76}.

The coupling of the wave vector modes in \gls{e1}{e2} explains the qualitative
trend that the correlators with larger 
$\alpha$--process amplitudes, $f^c_q$, relax
slower, i.e. have a larger $\tau_q$ \cite{Goetze91b,Fuchs92b,Franosch97}.
 Intriguingly, for a hard sphere liquid
the wave vector dependence of the dimension-less time scale $\tau^{\rm (f)}_q =
-r_s^2/(d^2 \ln{f^c_q})$, is rather close to the one of the actual
$\alpha$--time,
$\tau_q$, at least for intermediate wave vectors. Note that the comparison
shown 
in Fig. \ref{fig4} must be taken with a grain of salt, as the definition of
$\tau_q$ is not unique because of the stretching, i.e. non--exponentiality, of
the $\alpha$--process in MCT. Nevertheless, this semi--quantitative connection
of the $\alpha$--process amplitude to its time scale,
suggests a possible (partial) collapse of the $\Phi_q(t)$
for different $q$ at the same packing fraction onto a common curve given by:
\beq{e6}
\Phi_q(t) = \exp{\{- \frac{\Delta r^2(t)}{6\, d^2\, \tau^{\rm (f)}_q}\}}\; , 
\wer
 \tau^{\rm (f)}_q =\frac{-r_s^2}{d^2 \ln{f^c_q}}\; .
\eeq
Conceptually, $\Delta r^2(t)$ should be connected to the mean-squared
displacement of a colloid particle, to be denoted by $\delta r^2(t)$. 
From the
definition of $\tau^{\rm (f)}_q$, \gl{e6}, and the factorization property,
\gl{e4}, immediately follows that very close to the critical packing fractions
all rescaled curves intersect at the $\beta$--scaling time
$t_\varepsilon$. The connection of $f_q^c$ via $\tau^{\rm (f)}_q$ to the 
$\alpha$--relaxation time, see Fig. \ref{fig4},
then shepherds the correlators to stay close during the final relaxation step,
too.  Figure
\ref{fig5} shows representative scaled correlators for two packing fractions,
where the used wave vectors are marked in the inset. The correlators are drawn
for $\Phi_q(t)\ge 0.05$ in order to prevent overcrowding the figure. The
$q$--dependent stretching of the correlators causes  a noticeable spreading of
the  rescaled correlators  for long times. Considering Fig. \ref{fig4},
one also does not expect a data collapse for wave vectors outside the shown
$q$--range. Moreover, this scaling explicitly
violates the short time behavior of the intermediate scattering functions
which e. g. become diffusive for small wave vectors invalidating \gl{e6}. This
explains the spread of the curves in Fig. \ref{fig5} at short times.
 In Fig. \ref{fig5} also the
mean--squared  displacement from Ref. \cite{Fuchs98}
shifted as suggested by \gl{e6} is shown and lies
within the clatter of the curves. 

This ansatz, together with the known Gaussian
approximation to the self intermediate scattering function
\cite{hansen,Pusey91,Fuchs98},  gives a most
simplistic description of the coherent and incoherent density correlators of
the MCT. Nevertheless, the only point where \gl{e6}
asymptotically rigorously collapses all correlators is at
$\Phi_q(t_\varepsilon)=f^c_q$ 
because of the factorization property, \gl{e4}. Already in a vicinity of this  
point a spread of the curves exists because of  $h_q\ne h_{\rm
MSD}f^c_q/(6 d^2 \tau^{\rm (f)}_q)$,  which would follow from \gl{e6} 
and the known $\beta$--expansion, $\delta r^2(t) \dot{=}\,
r_{\rm sc}^2 - h_{\rm MSD} G(t)$ \cite{Goetze91b,Fuchs98}.

\subsection{Corrections}\label{sekt3b}

The discussion up to now has used the asymptotic formulae to lowest
orders in the separation parameter and thus might restrict the discussed
phenomena to close neighborhoods of the critical packing fraction $\phi_c$.  
The leading corrections in $\varepsilon$ to the asymptotic scaling laws of
Sect. \ref{sekt3a} have recently been discussed in detail for the present model
\cite{Franosch97,Fuchs98}, and in some cases allow to extend the range of
validity of the asymptotic expansions appreciably.  

The corrections to the $\beta$--scaling law, \gl{e4}, for the dynamics close to
$f^c_q$  are of the form:
\beq{e7}
\Phi_q(t) = f^c_q + h_q\, (\, G(t) + H(t) + K_q\, G^2(t) + 
\varepsilon \tilde{K}_q\, )
\; , 
\eeq
where the $K_q$ and $\tilde{K}_q$ 
 are wave vector dependent constants which follow from asymptotic solutions to
\gls{e1}{e2}. See \cite{Franosch97,Fuchs98} for the definitions and for the
correction function $H(t)$, which is of order ${\cal O}(\varepsilon)$.
The range of validity of the $\beta$--scaling law, \gl{e4}, is
thus found to be of order $\sqrt{\varepsilon}$, 
and to differ for  different wave vectors or observables. The 
$\beta$--region description of \gl{e7} extends the range of usefulness of the
MCT asymptotic expansion around the critical non--ergodicity plateau
appreciably as can be seen in Refs. \cite{Franosch97,Fuchs98}, 
and provides detailed few
parameter formulae for the density correlators which have already found use in
the data analysis of computer simulation studies
\cite{Sciortino97,Kaemmerer98,Bennemann99,Gleim99}. 
For the curves without HI
 of Fig. \ref{fig2}, \gl{e7} describes the  correlators in the 
$\beta$--window on a
10\% error level starting from the time scale $t\, \gtrapprox\, 0.1 d^2/D_0$ 
which was estimated in Sect. \ref{sekt3a} to be the range of domination of the
structural relaxation.  Thus \gl{e7} extends the asymptotic expansions 
of the structural relaxation almost
to the microscopic dynamics in this model. 

The $\alpha$--process has been the
focus of the recent DLS scattering studies \cite{segre96}
and, for wave vectors around the
peak,   describes the main portion of the decay of $\Phi_q(t)$. 
The range of validity of the $\alpha$--process--superposition
principle, \gl{e5}, is appreciably larger than the one of the $\beta$--scaling
law:
\beq{e8} 
\Phi_q(t) \to \tilde{\Phi}_q(t/t'_\varepsilon) + \varepsilon
\tilde{\Psi}_q(t/t'_\varepsilon) 
\fur \varepsilon\to0-
\;.
\eeq
The corrections in \gl{e8} are only of linear order in $\varepsilon$. Although
the complete form of $\tilde{\Psi}_q(\tilde{t})$ is not known yet, its
variation for short rescaled times can be deduced and can be argued to give the
dominant correction to \gl{e5} for not too large  $\varepsilon$:
$\tilde{\Psi}_q(\tilde{t} \to 0) = - h_q\, \tilde{B_1}\,
\tilde{t}^{-b}$, where the coefficient $\tilde{B_1}$ is of order unity. As this
term can grow without bounds, 
the dominant aspect of the leading corrections is to cause the correlators
$\Phi_q(t)$ to rise above the $\alpha$--master curves for times shorter than
the $\alpha$--relaxation time.  In this time window, the $\alpha$--master
curves follow von Schweidler's law, $\tilde{\Phi}_q(\tilde{t})-f_q^c= -
h_q\, \tilde{B}\, \tilde{t}^b$ (again with $\tilde{B}={\cal O}(1)$)
\cite{Goetze84,Goetze85}.  
As the $q$--dependence of the time scales of the
$\alpha$--correlators $\tilde{\Phi}_q(\tilde{t})$ can be estimated from
$\tilde{\Phi}_q(\tilde{t})\dot{=} f^c_q ( 1- (t/\tau_q^{\rm (vS)})^b)$, with
$\tau_q^{\rm (vS)}= (\tilde{B} f^c_q/h_q)^{1/b}\, t_\varepsilon'$
\cite{Goetze91b}, the short time
corrections can be rewritten in the time window $t_\varepsilon \ll t \ll
t_\varepsilon'$: 
\beq{e9}
\Phi_q(t) \dot{=} \; f^c_q\,  (\, 1 -  (t/\tau_q^{\rm (vS)})^b - \varepsilon \,
(\frac{t_\varepsilon'}{\tau_q^{\rm (vS)}})^{2b}\, 
(t/\tau_q^{\rm (vS)})^{-b}\, ) \; .
\eeq
Thus it is apparent that deviations from the asymptotic  $\alpha$--process
scaling law, \gl{e5}, are stronger for correlators with a shorter
$\alpha$--relaxation time or smaller $\alpha$--process amplitude; the second
connection arising  because of the relation between
$\tau_q$ and $f^c_q$, see Fig. \ref{fig4}.  

If at larger separations from the critical density the $\alpha$--relaxation
times are determined from the correlators $\Phi_q(t)$, then the corrections to
the $\alpha$--scaling law, \gl{e8}, may differently affect $\tau_q(\phi)$.  
This is caused by the inherent stretching in the $\alpha$--master curves 
\cite{Fuchs92b} and by
the time variation of the corrections $\tilde{\Psi}_q(\tilde{t})$. The dominant
short time variation of $\tilde{\Psi}_q(\tilde{t})$ leading to \gl{e9}  will 
affect $\tau_q(\phi)$ if a definition of the relaxation times is used, which
stresses the initial decay during the $\alpha$--process. 
A possible definition of $\tau_q(\phi)$ exhibiting this effect
is given by: $\Phi_q(t=\tau_q) = \frac 12 f^c_q$. Some results
are indicated in Fig. \ref{fig2}, where also the 
$\alpha$--process amplitudes, $f^c_q$, are shown in the inset. 
As Fig. \ref{fig6} shows, this definition of $\tau_q$ asymptotically 
gives almost identical
$q$--dependences as obtained from the $\tilde{\Phi}_q(\tilde{t})$
\cite{Fuchs92b}. 
  
Because of the rather large range of validity of the $\alpha$--scaling law,
\gl{e5}, for the intermediate scattering functions as explained by \gl{e8},
only very small deviations of $\tau_q(\phi)$ 
from the asymptotic wave vector dependence are seen in $\tau_q
(-\varepsilon)^\gamma$ for $n>3$. As expected from \gl{e9}, 
for $n=3$, which lies close to the limit of applicability of $\alpha$--scaling,
the relaxation times are relatively  longer and the largest (relative)
deviations appear for correlators with small $f^c_q$ or $\tilde{\tau}_q$. 
At this separation from the critical density, $\phi=0.9\, \phi_c$, already some
differences for the two models of the short time diffusion, with and without
HI, are noticeable in Fig. \ref{fig6}. As shown in the inset, the differences
can almost completely be incorporated into a packing fraction dependent shift
of the matching time $t_0$. If the time scales are
 normalized to unity for $q=q_p$,
  then  collapse can be achieved of the $\tau_q$ at $n=3$
except for the smallest wave vectors. 
Note that some finer aspects of the figure depend on the special choice how to
measure $\tau_q$. For example, the correlators without HI at $n=3$ and
for $q=q_p$ and $q=0.94 q_p$ (just below it) 
actually almost overlap and the apparent differences in $\tau_q$
arise solely from the  $f^c_q$--values entering the used definition. 

For even larger separations from the critical density, $n=2$ and $n=1$ in 
Fig. \ref{fig6}, clear differences of the long--time scales with and without HI
appear and can obviously not
be explained by structural relaxation, \gls{e1}{e2}, alone.
The diffusive particle motion on short time scales  causes the correlators for
small wave vectors to decay slower relative to the non--diffusive
$\alpha$--process. 

Often a diffusive behavior is assumed also for the structural relaxation and 
the relaxation times are converted to diffusion coefficients via $1/D_q=q^2
\tau_q$. Fig. \ref{fig7} shows so calculated $D_q$ normalized at $q=q_p$ in
order to eliminate the drift of $q_p$ with packing fraction; see
Fig. \ref{fig1}.
 Almost no deviations from the asymptotic variation as
follows from the $\alpha$--scaling law, \gl{e5}, can be be seen for $n>3$. 
Note that the non--diffusive character of the structural relaxation is hidden
in this representation.
For  larger separations and thus smaller packing fractions, a trend of the
long--time diffusion coefficients with hydrodynamic interactions (HI) to
approach the shape of the short time ones, $D_q^{\rm s. (HI)}$,
 can be recognized.

Considering Figs. \ref{fig6} and \ref{fig7}, one needs to keep in mind,
however, that differing methods to determine the final relaxation times or the
long--time diffusion coefficients would lead to somewhat
different $q$--dependences
because they would weigh the stretching of the $\alpha$--process
$\tilde{\Phi}_q(\tilde{t})$ and the leading corrections
$\tilde{\Psi}_q(\tilde{t})$ differently. The definitions chosen here allow to
explain the wave vector and packing fraction dependences  in $\tau_q$ and $D_q$
from known aspects, \gls{e8}{e9}, of the asymptotic expansions.  

\subsection{Visco--elastic properties}\label{sekt3z}
 
The time or frequency dependent shear modulus, $G_\eta$, 
of colloidal suspensions can be
defined as an autocorrelation function of  elements of the  stress
tensor and splits into three contributions \cite{Batchelor77,naegele98}. 
Whereas the first arises 
from the direct potential interactions of the particles and is familiar from
simple atomic liquids, the latter two contain effects of the HI and are
peculiar for colloidal particles immersed in a solvent. Only for the first
potential part there exist MCT expressions which are applicable close to the
glass transition at $\phi_c$ \cite{Bengtzelius84,Goetze91b,Fuchs92b}; 
however see Ref. \cite{naegele98} for lower
densities. Similarly as for $m_{q}(t)$ from \gl{e2}, $G_\eta(t)$ is given by
a quadratic mode--coupling functional in the $\Phi_q(t)$. 
Consistent with the neglect of the HI contributions to $G_\eta(t)$, solutions for
the $\Phi_q(t)$ are used which are calculated without HI, i. e. with the 
short--time diffusion coefficients $D^{\rm s. (B)}_q$.  
Figure \ref{fig8} shows the frequency dependent storage and loss shear
moduli for a number of densities \cite{Gdefinition}.  
As we consider the part of $G_\eta$ arising from potential colloidal interactions
only,  and thus cannot address the importance of HI at higher frequencies,
only results in the frequency window of structural relaxation are shown.
For low frequencies, the  viscosity $\eta$ can be obtained via
$G_\eta''(\omega\to0)\to\omega (\eta-\eta_\infty)$, where $\eta_\infty$ is the
high--frequency  shear viscosity which is caused by instantaneous solvent
interactions \cite{beenakker84,segre95b,Bergenholtz98b}.
We use the approximation $\eta_\infty=k_BT/(3\pi d D_0)$.
A plateau region in $G_\eta'(\omega)$ corresponds
to the $\beta$--scaling window, \gl{e4}, and indicates elastic behavior of the
colloidal suspension on intermediate time scales. In the non--ergodic states
above $\phi_c$, the colloidal system would be characterized by a finite elastic
shear modulus, $G_\eta\ge G^c_{\eta}$, where the value at the glass transition
follows from the $f^c_q$. The appropriate Fourier transforms of the
$\beta$--correlator describe the dynamics around this elastic plateau in
$G_\eta'$, and in the minimum region of $G_\eta''(\omega)$ between the transient
high frequency dynamics and the $\alpha$--relaxation peak, which sensitively
shifts  with separation from the critical density.

For a single colloidal particle in a continuum fluid the Stokes--Einstein
relation connects the particle diffusion coefficient and the solvent viscosity,
$\eta D^{\rm Self} = \frac{k_BT}{3\pi d}$. The self diffusion coefficient and
the mean--squared displacement at finite colloid densities 
can, within MCT, be obtained from the autocorrelation function of the
fluctuating forces which the single particle experiences from the colloidal
liquid \cite{Fuchs98}:
\beq{e10}
\delta r^2(t) + D^{\rm s. Self} \int_0^t dt' \, m^{\rm Self}(t-t') \, \delta r^2(t')
= 6 D^{\rm s. Self}\, t\; ,
\eeq
where $D^{\rm s. Self}$ is the short--time diffusion coefficient of the single
particle which, neglecting HI, is given by 
$D^{\rm s. Self} = D_0$ \cite{Pusey91}. 
The long--time self diffusion coefficient, $D^{\rm
Self}$,  follows from \gl{e10} in the Markovian limit, 
$D^{\rm s. Self}/ D^{\rm Self} = 1+ D^{\rm s. Self}\int_0^\infty dt\,
m^{\rm Self}(t)$ .
The memory function $m^{\rm Self}$ in MCT is given by another mode--coupling
functional.
 Thus a priori, within MCT one would expect connections or similarities of
$m^{\rm Self}(\omega)$ and $G_\eta(\omega)$ only because of the scaling laws.
In the $\beta$--scaling region, asymptotically both functions 
exhibit the same shape \cite{Goetze91b}, 
$G_\eta''(\omega)/h_{G_\eta}\to \chi''(\omega)$
and ${m^{\rm Self}}''(\omega)/h_{m^{\rm Self}}\to \chi''(\omega)$, where 
$\chi''(\omega)$ follows from the $\beta$--correlator $G(t)$ in \gl{e4}. 
It is included in Fig. \ref{fig8}. The $\alpha$--superposition principle,
\gl{e5}, states that the $\alpha$--relaxation peaks in both functions
asymptotically approach a  density independent shape and  shift in
parallel upon varying $\varepsilon$.
This $\alpha$--scale coupling
also immediately predicts the product $D^{\rm Self} \eta$ to approach a
constant asymptotically for $\phi\nearrow\phi_c$ \cite{Goetze91b}.
 Nevertheless, as for example the $\alpha$--peak positions need not coincide,
the close agreement of $G_\eta(t)$
and $m^{\rm Self}(t)$ in Fig. \ref{fig8}
over a wide window in time or frequency and covering a
substantial variation in packing fraction is somewhat surprising. Presumably it
arises, because, during the cooperative structural motion (cage effect), the
collective 
density correlators around the peak in $S_q$, i.e. on the length scale of the
average particle distance,  dominate the dynamics of small--$q$ MCT memory
functions.

\subsection{Comparison with experiments}\label{sekt3d}

The results of the MCT calculations of the previous sections, which partially
have been tested in DLS experiments aimed at the glass transition 
\cite{Megen91,Megen91b,Megen93,Megen93b,Megen94,Megen94b,Megen95,Megen98},
can also be used to discuss the recent experiments 
\cite{segre96,pusey97,segre97,Segre95,pusey97,mason95b,Mason95}
 at somewhat lower densities which were
mentioned in the introduction.
 Various other aspects of the
results and their possible experimental relevance have been presented in
\cite{Bengtzelius84,Barrat89,Goetze91b,Fuchs92b,Fuchs95,Franosch97,Fuchs98} and
will not be repeated here.  

As a first aspect, let us point out, that if the mean--squared displacement can
be measured and thus the connected memory function, $m^{\rm Self}$, then the
numerical results show that a close estimate of the potential part of the
shear modulus $G_\eta$ can be obtained. Even beyond the
connections predicted by the two asymptotic scaling laws, \gls{e4}{e5}, the
numerical results exhibited in Fig. \ref{fig8}, show that both functions are
closely related, presumably because both arise from  the cooperative cage
dynamics. This connection may be considered as a frequency dependent
generalization of the Stokes--Einstein relation and was assumed and tested in
the recent diffusive wave spectroscopy measurements of Mason and Weitz
\cite{mason95b}.  
In another study of the same authors \cite{Mason95}, they also observed that
the $\beta$--correlators from \gl{e4} provide a description of the
(directly measured) shear moduli spectra in an intermediate frequency window
consistent with the MCT description of the potential part of $G_\eta$. 

The $\alpha$--scale coupling predicts that the various relaxation times and
transport coefficients of a colloidal suspension close to the critical packing
fraction $\phi_c$ shift in parallel. For example, the prediction 
$\frac{2k_BT}{\pi d\eta D^{\rm Self}}=$ const. for $\phi\nearrow\phi_c$ follows
from \gl{e5}. Quantitatively, the ratio approaches 5.93 \cite{Fuchs92b}, 
see Fig. \ref{fig9},
a value very close to
the classical Stokes--Einstein prediction. Note however, that the conditions
required for the classical Stokes--Einstein relation to hold, clearly are
violated at packing fractions around the glass transition.  
A small but noticeable packing fraction dependence in $1/(D^{\rm Self}\eta)$
arises because of trivial density prefactors connecting the exhibited moduli
of Fig. \ref{fig8} with the transport coefficients. 
The $\alpha$--process corrections \gl{e8} and their  discussion in 
\gl{e9} suggest that the $\alpha$--scale coupling should hold well for 
$\alpha$--relaxation scales obtained at low frequencies.  
This is supported by the observation, that the Stokes--Einstein relation 
considered with
$\lim_{\omega\to0} \phi_c G_\eta''(\omega)/(\phi \omega)$  replacing $\eta$,
and ${m^{\rm Self}}''(\omega=0)$ replacing $1/D^{\rm Self}$,
considerably reduces its density dependence. Upon decreasing the
packing fraction to $\phi = 0.9\, \phi_c$ ($n=3$), where the $\alpha$--scaling
law looses validity, this ratio increases by 20 \% 
relative to the asymptotic value, whereas the actual 
Stokes--Einstein ration increases by 29 \%. Even larger
density dependences can be expected if the long--time diffusion 
coefficients are obtained in time or frequency windows where the dominant
corrections to the $\alpha$--process, see \gl{e9}, appreciably 
increase the relaxation times of the correlators with shorter
asymptotic  $\alpha$--relaxation times. 
This effect is apparent in Fig. \ref{fig9},  where the  wave vectors
away from the peak position in $S_q$ show an increase in the relaxation times
relative to the asymptotic $\alpha$--scale prediction,
which on the other hand
  holds rather well for $q=q_p$. At $q=1.17 q_p$ where $S_q(\phi_c)=0.90$
a 65 \% increase is seen at $n=3$, whereas at $q=q_p$ the Stokes--Einstein
ratio changes only by 11 \%. 

The results concerning the generalized Stokes--Einstein relation, which are 
presented in Fig. \ref{fig9}, and their explanations using
\gls{e8}{e9} rest on the simplifications caused by the bifurcation
singularity in the MCT equations and by the entailing asymptotic
expansions. As discussed, the range of validity of the $\alpha$--scale coupling
does not  appreciably exceed $\varepsilon\approx-0.1$ 
for the studied models of hard sphere like
colloidal suspensions, and thus microscopic corrections
to the MCT description of the structural relaxation need to be incorporated in
principle beyond this distance to $\phi_c$. Thus it is consistent with MCT
that for packing fractions well below $\phi_c$ the coupling of the time scales
may continue in one system (hard spheres) but not in another one (charged
spheres) \cite{johanprl}. Also the estimate for the range of
packing fractions where $\alpha$--scaling should hold can be expected to be
model dependent, as the $\alpha$--master functions, \gl{e5}, and their
corrections, \gl{e8}, depend on the fluid structure.

Approximate
expressions might be useful to describe the qualitative trends in the
intermediate scattering functions, $\Phi_q(t)$, in a wider context, 
in the same way as
the Gaussian approximation \cite{hansen} is useful for the
self intermediate scattering functions, $\Phi^s_q(t)$. The Gaussian
approximation was compared  to the MCT results in \cite{Fuchs98}.
  Recently Segr{\`e} and Pusey 
were lead by their DLS scattering data to propose such a formula \cite{segre96}
and they observed partial collapse of their rescaled data.  

In \gl{e6} the possibility to collapse the intermediate scattering functions
onto a common curve is studied using the dimensionless
time scale $\tau^{\rm (f)}_q$. This is suggested by the finding that 
the $\alpha$--relaxation amplitudes and relaxation times asymptotically 
are connected via $\tau^{\rm (f)}_q$; see  Fig. \ref{fig4}. 
The qualitative connection was expected \cite{Goetze91b,Fuchs92b} 
but the quantitative closeness surprises and may be peculiar to
the hard sphere system.
Satisfactory data collapse using \gl{e6}, see Fig. \ref{fig5},
is possible with deviations at long times because of the non--universality of
the  
$\alpha$--process, and at short times, because \gl{e6} violates the short--time
diffusive motion of colloidal suspensions. 
This reiterates that within MCT there is no connection of the obtained
long--time  diffusion coefficients, which follow from \gls{e1}{e2},
to the short--time ones \cite{Franosch98}. 
The only effect of the latter could be a
 shift in the time scale $t_0$, which
matches the structural relaxation to the microscopic transient. 

Similar shapes, however, of the long--
and the short--time diffusion coefficients with HI were
 observed by Segr{\`e} and Pusey in the recent DLS experiments 
on colloid fluids below and close to the glass transition
\cite{segre96,segre95b}. 
This similarity of the short-- and long--time diffusion coefficients
 suggests to collapse the  intermediate scattering functions
with the assumption \cite{segre96}:\newline
 $\Phi_q(t) = \exp{\{ - (q^2/6) (D^{\rm
s.}_q/D^{\rm s. Self}) 
\delta r^2(t) \} }$, \newline
which becomes exact for short times.
 In their experiments,  Segr{\`e} and Pusey  observed  data collapse for
 wave vectors 
starting from somewhat below $q_p$ to  the  position of the second maximum in
$S_q$.  
Figure \ref{fig10} shows the solutions of the MCT equations with $D^{\rm
s. (HI)}_q$ appropriately rescaled. 
Reasonable collapse of the curves onto a common one, which also is well
represented  
by the mean--squared displacement, is observed
in a similar wave vector range as in the experiments.
 For short times all curves coincide rigorously.
For small wave vectors the non--diffusive character of the $\alpha$--process 
however leads to strong deviations for longer times. This trend also is
present in the experimental data. 
The non--diffusive structural relaxation disagrees with the
 assumed  diffusive scaling  of the density correlators and thus cannot be
 rationalized 
with considerations of the short--time expansions following de Gennes.
Unavoidable polydispersity effects in the experimental data could lead to
additional deviations from the proposed scaling for small $q$,
 but no qualitative differences
for samples of different polydispersities were reported in
\cite{segre96,segre97}. Polydispersity effects could be incorporated into the
present MCT following the work for charged colloids in Ref. \cite{Baur96}.
The partial collapse of the data for intermediate and long times results from
the connections 
of the $\alpha$--process amplitudes to the time scales discussed in context
with the ansatz of \gl{e6}. Note that this connection may not
be quantitatively 
satisfied  as well in other colloidal systems, like e. g. charged
colloidal particles \cite{haertl95}. Thus the approximate scaling may
hold less well in other systems.
Differently from \gl{e6},  the scaling with the short--time
diffusion coefficients 
does not rigorously collapse the data at a longer time.
But the close similarity of $\tau^{\rm (f)}_q$ and $D^{\rm s. (HI)}_q$ explains
that   collapse for short times and approximate collapse for longer times is
achieved. 

\section{Conclusions}\label{sekt4}

In the idealized MCT, the long--time dynamics of colloidal liquids is dominated
by the structural relaxation. Asymptotic expansions close to the critical
packing fraction capture the qualitative aspects of the structural
relaxation. In this contribution it is shown that the theoretical results also
rationalize some recent experimental findings for larger separations from the
critical density. Corrections to the coupling of the $\alpha$--relaxation times
are wave vector dependent as seen in tests of generalized Stokes--Einstein
ratios \cite{Segre95}. The latter can be generalized to finite frequencies as
observed in light scattering experiments \cite{mason95b}, if 
the potential contribution to the shear modulus is considered. The 
tight coupling  of the collective density fluctuations  as captured in the
scaling observed by  Segr{\`e} and Pusey \cite{segre96} on the one hand
supports the existence of an $\alpha$--scaling law as predicted by MCT, and on
the other hand requires the quantitative connection of the $\alpha$--process
amplitudes and relaxation times obtained here for hard sphere like colloidal
particles.  The failure of the scaling for small wave vectors seen in the
experiments is predicted by MCT and highlights that the structural relaxation
cannot be understood from short--time expansions. Such an approach, often
referred to as ``de Gennes narrowing'' concept, would suppose  diffusive
colloidal dynamics for small wave vectors in disagreement with MCT and
experiment \cite{segre96}.  As the  structural relaxation is
determined by the equilibrium structure factor only, hydrodynamic interactions
affecting the short--time and transient dynamics can be incorporated into the
MCT without changing the long-time predictions.

\acknowledgments{
Valuable discussions with Dr. A. Latz,
Dr. P. N.  Segr{\`e} and Prof. W. G{\"o}tze are gratefully acknowledged.
This work was supported by the Deutsche
Forschungsgemeinschaft under Grant No. Fu 309/2.}
\bibliography{../main}

\begin{thebibliography}{10}

\bibitem{Pusey91}
P.~N. Pusey,  in {\em Liquids, Freezing and Glass Transition}, edited by J.-P.
  Hansen, D. Levesque, and J. Zinn-Justin (North-Holland, Amsterdam, 1991), p.\
  763.

\bibitem{russel}
W.~B. Russel, D.~A. Saville, and W.~R. Schowalter, {\em Colloidal Dispersions}
  (Cambridge UniversityPress, New York, 1989).

\bibitem{Megen91}
W. van Megen and P.~N. Pusey, Phys. Rev. A {\bf 43},  5429  (1991).

\bibitem{Megen91b}
W. van Megen, S.~M. Underwood, and P.~N. Pusey, Phys. Rev. Lett. {\bf 67},
  1586  (1991).

\bibitem{Megen93}
W. van Megen and S.~M. Underwood, Phys. Rev. Lett. {\bf 70},  2766  (1993).

\bibitem{Megen93b}
W. van Megen and S.~M. Underwood, Phys. Rev. E {\bf 47},  248  (1993).

\bibitem{Megen94}
W. van Megen and S. Underwood, Phys. Rev. Lett. {\bf 72},  1773  (1994).

\bibitem{Megen94b}
W. van Megen and S.~M. Underwood, Phys. Rev. E {\bf 49},  4206  (1994).

\bibitem{Megen95}
W. van Megen, Transp. Theory Stat. Phys. {\bf 24},  1017  (1995).

\bibitem{Megen98}
W. van Megen, T.~C. Mortensen, J. M{\"u}ller, and S.~R. Williams, Phys. Rev. E
  {\bf 58},  6073  (1998).

\bibitem{Goetze91}
W. G{\"o}tze and L. Sj{\"o}gren, Phys. Rev. A {\bf 43},  5442  (1991).

\bibitem{Goetze91b}
W. G{\"o}tze,  in {\em Liquids, Freezing and Glass Transition}, edited by J.-P.
  Hansen, D. Levesque, and J. Zinn-Justin (North-Holland, Amsterdam, 1991), p.\
  287.

\bibitem{gs}
W. G\"otze and L. Sj\"ogren, Rep. Prog. Phys. {\bf 55},  241  (1992).

\bibitem{Bartsch93b}
E. Bartsch, V. Frenz, S. M{\"o}ller, and H. Sillescu, Physica {\bf A 201},  363
   (1993).

\bibitem{Bartsch95c}
E. Bartsch, Transp. Theory Stat. Phys. {\bf 24},  1125  (1995).

\bibitem{haertl95}
W. H{\"a}rtl, H. Versmold, and X. Zhang-Heider, J. Chem. Phys. {\bf 102},  6613
   (1995).

\bibitem{gang}
H. Gang, A.~H. Krall, H.~Z. Cummins, and D.~A. Weitz, Phys. Rev. Lett. {\bf
  59},  715  (1999).

\bibitem{segre96}
P.~N. Segr{\`e} and P.~N. Pusey, Phys. Rev. Lett. {\bf 77},  771  (1996).

\bibitem{pusey97}
P.~N. Pusey, P.~N. Segr{\`e}, O.~P. Behrend, S.~P. Meeker, and W.~C.~K. Poon,
  Physica {\bf A 235},  1  (1997).

\bibitem{segre97}
P.~N. Segr{\`e} and P.~N. Pusey, Physica {\bf A 235},  9  (1997).

\bibitem{Segre95}
P.~N. Segr{\`e}, S.~P. Meeker, P.~N. Pusey, and W.~C.~K. Poon, Phys. Rev. Lett.
  {\bf 75},  958  (1995).

\bibitem{mason95b}
T.~G. Mason and D.~A. Weitz, Phys. Rev. Lett. {\bf 74},  1250  (1995).

\bibitem{Mason95}
T.~G. Mason and D.~A. Weitz, Phys. Rev. Lett. {\bf 75},  2770  (1995).

\bibitem{johanprl}
A.~J. Banchio, J. Bergenholtz, and G. N\"agele, Phys. Rev. Lett. {\bf 82},
  1792  (1999).

\bibitem{Leutheusser84}
E. Leutheusser, Phys. Rev. A {\bf 29},  2765  (1984).

\bibitem{Bengtzelius84}
U. Bengtzelius, W. G{\"o}tze, and A. Sj{\"o}lander, J. Phys. C {\bf 17},  5915
  (1984).

\bibitem{Goetze95}
W. G{\"o}tze and L. Sj{\"o}gren, Transp. Theory Stat. Phys. {\bf 24},  801
  (1995).

\bibitem{Goetze84}
W. G{\"o}tze, Z. Phys. B {\bf 56},  139  (1984).

\bibitem{Goetze85}
W. G{\"o}tze, Z. Phys. B {\bf 60},  195  (1985).

\bibitem{Das86}
S.~P. Das and G.~F. Mazenko, Phys. Rev. A {\bf 34},  2265  (1986).

\bibitem{Goetze87}
W. G{\"o}tze and L. Sj{\"o}gren, Z. Phys. B {\bf 65},  415  (1987).

\bibitem{Fuchs92b}
M. Fuchs, I. Hofacker, and A. Latz, Phys. Rev. A {\bf 45},  898  (1992).

\bibitem{Franosch97}
T. Franosch, M. Fuchs, W. G{\"o}tze, M.~R. Mayr, and A.~P. Singh, Phys. Rev. E
  {\bf 55},  7153  (1997).

\bibitem{Fuchs98}
M. Fuchs, W. G{\"o}tze, and M.~R. Mayr, Phys. Rev. E {\bf 58},  3384  (1998).

\bibitem{Feynmann56}
R.~P. Feynmann and M. Cohen, Phys. Rev. {\bf 102},  1189  (1956).

\bibitem{Feynmann57}
R.~P. Feynmann and M. Cohen, Phys. Rev. {\bf 107},  13  (1957).

\bibitem{Goetze76}
W. G{\"o}tze and M. L{\"u}cke, Phys. Rev. B {\bf 13},  3825  (1976).

\bibitem{hansen}
J.~P. Hansen and I.~R. McDonald, {\em Theory of Simple Liquids} (Academic
  Press, London, 1986).

\bibitem{Sjoegren80}
L. Sj{\"o}gren, Phys. Rev. A {\bf 22},  2866  (1980).

\bibitem{Fuchs95}
M. Fuchs, Transp. Theory Stat. Phys. {\bf 24},  855  (1995).

\bibitem{degennes59}
P.~G. de~Gennes, Physica {\bf 25},  825  (1959).

\bibitem{beenakker83}
C.~W.~J. Beenakker and P. Mazur, Physica {\bf 120 A},  388  (1983).

\bibitem{beenakker84}
C.~W.~J. Beenakker and P. Mazur, Physica {\bf 126 A},  349  (1984).

\bibitem{segre95b}
P.~N. Segr{\`e}, O.~P. Behrend, and P.~N. Pusey, Phys. Rev. E {\bf 52},  5070
  (1995).

\bibitem{naegele97}
G. N\"agele and P. Baur, Europhys. Lett {\bf 38},  557  (1997).

\bibitem{Cohen91}
E.~G.~D. Cohen and I.~M. de~Schepper, J. Stat. Phys. {\bf 63},  241  (1991).

\bibitem{Cohen98}
E.~G. D. C.~R. Verberg, , and I.~M. de~Schepper, Physica {\bf A 251},  251
  (1998).

\bibitem{toku1}
M. Tokuyama and I. Oppenheim, Physica {\bf A 216},  85  (1995).

\bibitem{toku2}
M. Tokuyama, Physica {\bf A 229},  36  (1996).

\bibitem{toku3}
M. Tokuyama, Y. Enomoto, and I. Oppenheim, Phys. Rev. E {\bf 56},  2302
  (1997).

\bibitem{Fabbian99}
L. Fabbian, W. G\"otze, F. Sciortino, P. Tartaglia, and F. Thiery, Phys. Rev. E
  {\bf 59},  R1347  (1999).

\bibitem{Bergenholtz99}
J. Bergenholtz and M. Fuchs, Phys. Rev. E {\bf 59},  5706  (1999).

\bibitem{Goetze87b}
W. G{\"o}tze,  in {\em Amorphous and Liquid Materials}, {\em NATO ASI Series},
  edited by E. L{\"u}scher, G. Fritsch, and G. Jacucci (Martinus Nijhoff Pub.,
  Dordrecht, 1987), p.\ 34.

\bibitem{Barrat89}
J.-L. Barrat, W. G{\"o}tze, and A. Latz, J. Phys.: Condens. Matter {\bf 1},
  7163  (1989).

\bibitem{Franosch98}
T. Franosch, W. G{\"o}tze, M.~R. Mayr, and A.~P. Singh, J. Non-Cryst. Solids
  {\bf 235--237},  71  (1998).

\bibitem{taudef1}
As the MCT predicts a non--exponential $\alpha$--relaxtion process, the
  definition of $\tau_q$ is not unique \cite{Fuchs92b}. A useful definition is
  $(d/d \ln \tilde t)^2 \tilde{\Phi}_q(\tilde{t}) =0$ for
  $\tilde{t}=\tilde{\tau}_q$, which will be used in section \ref{sekt3a}.

\bibitem{Mountain66}
R.~D. Mountain, J. Res. Nat. Bur. Standards {\bf 70A},  207  (1966).

\bibitem{Sciortino97}
F. Sciortino, L. Fabbian, S.-H. Chen, and P. Tartaglia, Phys. Rev. E {\bf 56},
  5397  (1997).

\bibitem{Kaemmerer98}
S. K{\"a}mmerer, W. Kob, and R. Schilling, Phys. Rev. E {\bf 58},  2131
  (1998).

\bibitem{Bennemann99}
C. Bennemann, J. Baschnagel, and W. Paul, J. Phys.: Condens. Matter {\bf XX},
  XXXX  (1999).

\bibitem{Gleim99}
T. Gleim and W. Kob, Phys. Rev. Lett. {\bf XX},  XXXX  (1999).

\bibitem{Batchelor77}
G.~K. Batchelor, J. Fluid Mech {\bf 83},  97  (1977).

\bibitem{naegele98}
G. N\"agele and J. Bergenholtz, J. Chem. Phys. {\bf 108},  9893  (1998).

\bibitem{Gdefinition}
The notation $G_s''(\omega)+iG_s'(\omega)=\omega \int_0^\infty dt e^{i\omega t}
  G_s(t)$ is used.

\bibitem{Bergenholtz98b}
J. Bergenholtz, F.~M. Horn, W. Richtering, and N.~W. N.~J. Wagner, Phys. Rev. E
  {\bf 58},  R4088  (1998), and references therein.

\bibitem{Baur96}
P. Baur, G. N\"agele, and R. Klein, Phys. Rev. E {\bf 53},  6224  (1996).

\end{thebibliography}
\bibliographystyle{prsty}
\end{multicols}

\clearpage
\begin{figure}[ht]
\caption[Figur eins]{\label{fig1}}
Inverse short--time diffusion coefficients without hydrodynamic interactions
(HI), $D^{\rm s. (B)}_q$ (left scale), and with HI, $D^{\rm s. (HI)}_q$
(dot--dashed curve, right scale), versus rescaled wave vector.
Without HI, the density variation is determined by the structure factor,
$D_0/D^{\rm s. (B)}_q=S_q$, and is recorded for $\phi=\phi_c (1-10^{-n/3})$
with $n=1,2,3$ (thin dotted, short and
long dashed line) and $\phi_c=0.516\ldots$ \protect\cite{Franosch97};
 for $n\ge$6, the $S_q$ (almost) collapse onto the bold
solid
line. Due to the rough modeling, the shape of  $D^{\rm s. (HI)}_{q/q_p}$ is
not varied with density.   The inset shows the density dependence of the peak
position $q_p$ for the considered densities corresponding to $n=$ 1,2,3,6,9,12.
\end{figure}
\begin{figure}[ht]
\caption[Figur eins]{\label{fig2}}
Normalized intermediate scattering functions $\Phi_q(t)$ versus time
for five wave vectors indicated by symbols in the inset of part (b).
In (a) and (b), the height of the circles equals $f^c_q/2$ and the height of
the diamonds gives $f^c_q$. In (a) the packing   fraction is $\phi = 0.999\,
\phi_c$, $n=9$, and $\phi = 0.9\, \phi_c$, $n=3$, in (b). Dashed lines
marked with circles result from calculations without HI and solid lines
marked with diamonds from the ones with HI.  The inset shows the amplitude
$f^c_q$ of the final $\alpha$--relaxation process. The circles in (a) and
(b) indicate the relaxation times estimated from $\Phi_q(t=\tau_q)=\frac 12
f^c_q$. 
\end{figure}
\begin{figure}[ht]
\caption[Figur eins]{\label{fig4}}
Dimensionless time scales $\tau_q^{\rm (f)}$ (full circles and left axis) 
resulting from $f^c_q$, see Eq.
(\protect\ref{e6}), versus wave vector and compared to the rescaled
$\alpha$--relaxation times (open circles and right axis) 
\protect\cite{taudef1}. The lines through the points
indicate the corresponding results for the Verlet--Weis $S_q$ from
\protect\cite{Fuchs92b}. The short dashed curve shows $S_q/(qd)^2$
appropriately shifted to match at $q_p$.  
\end{figure}
\begin{figure}[ht]
\caption[Figur eins]{\label{fig5}}
Intermediate scattering functions versus time replotted as suggested by Eq.
(\protect\ref{e6}) with $\tau_q^{\rm (f)}$ taken from Fig. \protect\ref{fig4};
an arbitrarily chosen factor enlarges the vertical scale, and the part
$\Phi_q(t)>0.05$ lies in the window for all but one
correlator. Fig. (a) presents
results for a density close to $\phi_c$, $\phi=0.999\, \phi_c$ ($n=9$),
 whereas Fig. (b) corresponds to $\phi=0.9\, \phi_c$ ($n=3$).  
The full solid lines correspond to the mean--squared
displacement $\delta r^2(t)$ from Ref. \protect\cite{Fuchs98} 
scaled accordingly. The wave vectors of the exhibited $\Phi_q(t)$  are marked
in the inset of part (b). It shows  $\tau_q^{\rm (f)}/ \tau_{q_p}^{\rm (f)}$
(symbols and solid line) and the $\alpha$--relaxation times
$\tau_q/\tau_{q_p}$ (dashed line) from Fig. \protect\ref{fig4}. 
\end{figure}
\begin{figure}[ht]
\caption[Figur eins]{\label{fig6}}
Reduced $\alpha$--relaxation times in units of $d^2/160D_0$
for various packing fractions and defined 
by $\Phi_q(t=\tau_q)=f^c_q/2$. The curves are 
 shifted according to the $\alpha$--scaling law, Eq. 
\protect\ref{e5} with $\gamma=2.46$ \protect\cite{Franosch97}, and plotted
versus wave vector. The thin overlapping lines are for
both models at $n=6$, $9$ and $12$. The small circles repeat the rescaled
Verlet--Weis  result from Fig. \protect\ref{fig4}.
The bold dashed line results for the model without HI at $n=3$ 
(left scale), and the bold solid line for the one with HI( right scale) at
$n=3$. The inset repeats the data for $n=3$ shifted to unity at 
$q_p$ and also includes curves for $n=1$ and $2$ (thin); line styles as in the
main  part. 
\end{figure}
\begin{figure}[ht]
\caption[Figur eins]{\label{fig7}}
Long--time diffusion coefficients calculated from the times in Fig.
\protect\ref{fig6} and normalized to unity at $q_p$ plotted versus rescaled
wave vector. For $n\ge 6$ the thin solid lines overlap and indicate the
asymptotic   result. Results without HI (bold long--dashed line) and with  HI
(bold solid line) at  $n=1$  are shown and compared to the rescaled short--time
diffusion 
coefficients at $n=1$,  $D^{\rm s. (B)}$ (thin short dashes) and $D^{\rm
s. (HI)}$ (thin dot--dashed line). 
\end{figure}
\begin{figure}[ht]
\caption[Figur eins]{\label{fig8}}
Loss shear modulus, $G_\eta''(\omega)$, 
solid lines and left scale, compared to the
self particle memory function, ${m^{\rm Self}}''(\omega)$ dashed line and right
scale,  for the separations from the critical packing fraction corresponding to
$n=$ 1, 2, 3, 6 and 9 as labeled. The inset shows the storage shear modulus
for the same packing fractions. The dashed--dotted curves in both cases
indicate the appropriate 
Fourier transforms of the $\beta$--correlator from Eq. \protect\ref{e4}. 
\end{figure}
\begin{figure}[ht]
\caption[Figur eins]{\label{fig9}}
Packing fraction dependent generalized Stokes--Einstein relations calculated
from the  viscosity $\eta$ and the
 long--time diffusion coefficients from Fig. \protect\ref{fig7} 
are shown for various
wave vectors indicated in the inset, which shows $S_q(\phi_c)$. Solid squares
use the long--time self diffusion coefficients. The open symbols show the
corresponding 
ratio using
$\lim_{\omega\to0} n_c G_\eta''(\omega)/(n \omega)$  instead of $\eta$,
and ${m^{\rm self}}''(\omega=0)$ instead of $1/D^{\rm Self}$. 
\end{figure}
\begin{figure}[ht]
\caption[Figur eins]{\label{fig10}}
Intermediate scattering functions with HI effects 
rescaled using the short--time diffusion coefficients, $D^{\rm s. (HI)}_q$ from
Fig. \protect\ref{fig1}, according to Ref. \protect\cite{segre96}; 
a factor enlarges the vertical scale, 
and the part $\Phi_q(t)>0.05$ lies in the window for all but three
correlators. Fig. (a) presents
Part (a) corresponds to a packing fraction $\phi=0.999\, \phi_c$ ($n=9$)
close to the critical density, whereas part (b) corresponds to a
 larger separation, $\phi=0.9\, \phi_c$ ($n=3$). The thick solid lines give 
the mean--squared displacement.
The other curves belong to the wave vectors indicated in the inset.
It shows the  normalized times  the short--time decay times, 
$\tau^{\rm s. (HI)}_q =1/(q^2D^{\rm s. (HI)}_q)$ (symbols and solid line) 
 and the $\alpha$--relaxation times
$\tau_q/\tau_{q_p}$ (dashed line) from Fig. \protect\ref{fig4}. 
\end{figure}

\end{document}